\documentclass[12pt]{iopart}

\usepackage{graphicx}
\usepackage{amssymb}
\usepackage{epstopdf}
\usepackage{color}
\usepackage{graphicx}
\usepackage{pdfpages}
\usepackage{ulem}
\begin{document}

\title{
Regular decoupling sector and exterior solutions in the context of MGD}

\author{Ernesto Contreras}
\address{Departamento de F\'isica, Colegio de Ciencias e Ingenier\'ia, Universidad San Francisco de Quito, Quito, Ecuador\\}
\ead{econtreras@usfq.edu.ec}
\author{Francisco Tello-Ortiz}
\address{Departamento de F\'isica, Facultad de ciencias b\'asicas, Universidad de Antofagasta, Casilla 170, Antofagasta, Chile}
\ead{francisco.tello@ua.cl}
\author{S. K. Maurya}
\address{Department of Mathematics and Physical Science, College of Arts and Science, University of Nizwa, Nizwa, Sultanate of Oman}
\ead{sunil@unizwa.edu.om}

\vspace{10pt}
\begin{indented}
\item[]August 2017
\end{indented}

\begin{abstract}
We implement the Gravitational Decoupling  through the Minimal Geometric Deformation method
and explore its effect on exterior solutions by imposing a regularity condition in the Tolman–Oppenheimer–Volkoff equation of the decoupling sector. We obtain that the decoupling function can be expressed formally in terms of an integral involving the $g_{tt}$ component of the metric of the seed solution. As a particular example, we implement the method by using the Schwarzschild exterior as a seed and we obtain that the asymptotic behavior of the extended geometry corresponds to a manifold with constant curvature.

\end{abstract}

%
%
%
%
%

\section{Introduction}\label{intro}
In search of new solutions of Einstein's field equations, we can try to extend well--known geometries by adding extra sources and then interpreting such additional matter sector in an appropriate manner. However, given the non--linearity of the Einstein field equations, it is clear that this strategy complicates the system of differential equations involved. In this respect, the first simple, systematic and direct way  of decoupling gravitational sources has been introduced in General Relativity \cite{ovalle2017,ovalleplb} in the framework of the Minimal Geometric Deformation (MGD) approach
\cite{ovalle2008,
	ovalle2009,ovalle2010,casadio2012,ovalle2013,ovalle2013a,
	casadio2014,casadio2015,ovalle2015,casadio2015b,
	ovalle2016, cavalcanti2016,casadio2016a,ovalle2017,
	rocha2017a,rocha2017b,casadio2017a,ovalle2018,ovalle2018bis,
	estrada2018,ovalle2018a,lasheras2018,gabbanelli2018,sharif2018,sharif2018a,sharif2108b,
	fernandez2018,fernandez2018b,
	contreras2018, estrada, contreras2018a,morales,tello18,
	rincon2018,ovalleplb,contreras2018c,contreras2019,contreras2019a,tello2019,	contrerasextended,tello2019a,lh2019,estrada2019,gabbanelli2019,ovalle2019a,sudipta2019,victorNS,linares2019,
	leon2019,casadio19,rincon19,Maurya1,jorgeLibro,Abellan1,Abellan2,roldao1,roldao2,newton19,tello20,tello2019b,o1,o2,o3,milkowski,sharifito,sharifito1,darocha,SK} which has opened a wide range of  possibilities	
to obtain new solutions of Einstein's equations. The method can be described as follows. Suppose that the energy--momentum tensor sustaining certain geometry is given by the superposition of a perfect fluid and some scalar, vector or tensor field. Now, after introducing a geometric deformation in the $g^{rr}$ component of a line element describing a static and spherically symmetric space--time,
Einstein's equations get decoupled in two sets of differential equations (one for each source involved)  and, as a consequence, the final solution is simply a superposition of the results obtained for each set separately. The method has been successfully implemented to obtain anisotropic like--Tolman IV solutions \cite{ovalle2017,ovalle2018}, 
anisotropic like--Tolman VII solutions \cite{sudipta2019} and models for neutron stars \cite{estrada2018,gabbanelli2018,morales,tello18,
tello2019,victorNS}, for example. Otherwise, in the context of modified theories of gravitation, the method has served  to obtain new solutions in $f(\mathcal{G})$ gravity \cite{sharif2018a}, Lovelock \cite{estrada2019}, $f(R,T)$ \cite{tello2019a}, Rastall \cite{tello2019b} and recently black holes and interior solutions in the context of braneworld \cite{leon2019}.

In other contexts, MGD has been used to extend black holes (BH) in $3+1$ and $2+1$ dimensional space--times \cite{ovalle2018a,contreras2018a,contreras2018c,rincon19}. For example, in
\cite{ovalle2018a} it was found the first extension of Schwarzschild BH resulting in a solution with extra critical points and, in some cases, naked singularities (in this case the solution can be considered as 
the exterior of a central object with radius greater than the point where the solution becomes singular). Besides, in \cite{rincon19} MGD was applied to extend the well known 
Reissner-Nordstr\"{o}m background, and it
was found that, after an elaborated analysis of the free parameters, the extended solution is free of extra singularities and satisfies all the energy conditions. 

It is worth mentioning that in all the extensions of exterior solutions mentioned above, the strategy is to impose a suitable equation of state in the decoupling sector such barotropic, polytropic or linear combinations of the component of the energy--momentum tensor. In this work, instead of following the standard procedure, we propose an alternative strategy
in which a
 regularity condition is imposed in the anisotropy induced by the decoupling sector in the framework of MGD, in order to ensure an acceptable behaviour of the Tolman–-Oppenheimer-–Volkoff (TOV) equation. As we shall see later, we base our study in the pioneering works of Bowers and Liang \cite{bowers} and Cosenza 
{\it et al}
\cite{cosenza1,cosenza2} where the same condition was applied to avoid the apparition of singularities in the anisotropy function in the context of stellar interiors. 

This work is organized as follows. In the next section we review the main aspects of MGD. In section \ref{new} we study the regularity condition in the decoupling sector and
illustrate the method using Schwarzschild's solution as a seed. Finally, the last section is devoted to final remarks.
Throughout the article we employ relativistic geometrized units where $G=c=1$ thus the overall constant $\kappa^{2}$ is equal to $8\pi$ and the mostly positive signature $\{-,+,+,+\}$.

\section{Minimal Geometric Deformation}\label{MGD}
This section is devoted to review the main aspects  of MGD in a self content way. Let us start by considering the  action
\begin{equation}\label{eq1}
S=S_{EH}+S_{M},    
\end{equation}
where 
\begin{equation}\label{eq2}
S_{E-H}=\frac{1}{2\kappa}\int \sqrt{-g}Rd^{4}x,    
\end{equation}
corresponds to the Einstein--Hilbert (EH) term
and 
\begin{equation}\label{eq3}
S_{M}=\int \sqrt{-g}\mathcal{L}_{M}d^{4}x,
\end{equation}
encodes the information of the matter content through
the matter Lagrangian $\mathcal{L}_{M}$. So, let us write $\mathcal{L}_{M}$ as
\begin{equation}\label{eq4}
\mathcal{L}_{M}=\mathcal{L}_{\bar{M}}+\alpha\mathcal{L}_{X},
\end{equation}
where $\mathcal{L}_{\bar{M}}$ is associated with
isotropic, anisotropic or charged fluids and $\mathcal{L}_{X}$ encipher the information of a new matter field. Now, taking variations respect to the inverse metric $\delta g^{\mu\nu}$ in (\ref{eq1}), we arrive to
\begin{equation}\label{eq5}
\frac{\delta S}{\delta g^{\mu\nu}}=0 \Rightarrow G_{\mu\nu}\equiv R_{\mu\nu}-\frac{R}{2}g_{\mu\nu}=\kappa^{2} T^{(tot)}_{\mu\nu},
\end{equation}
where $G_{\mu\nu}$ is the Einstein's tensor and $T^{(tot)}_{\mu\nu}$ is given by
\begin{equation}\label{eq6}
T^{(tot)}_{\mu\nu}=\underbrace{-2\frac{\delta\mathcal{L}_{\bar{M}}}{\delta g^{\mu\nu}}+g_{\mu\nu}\mathcal{L}_{\bar{M}}}_{T_{\mu\nu}}+\alpha\underbrace{\left(-2\frac{\delta\mathcal{L}_{X}}{\delta g^{\mu\nu}}+g_{\mu\nu}\mathcal{L}_{X}\right)}_{\theta_{\mu\nu}}.
\end{equation}
Then,
\begin{equation}\label{total}
T_{\mu\nu}^{(tot)}=T_{\mu\nu}+\alpha\theta_{\mu\nu},
\end{equation}
being $\alpha$ a 
Lorentz and general coordinate
invariant free parameter 
which measure the strength of the 
new matter sector. Finally, Einstein's equations can be written as
\begin{eqnarray}\label{efq}
R_{\mu\nu}-\frac{1}{2}R g_{\mu\nu}=T_{\mu\nu}+\alpha\theta_{\mu\nu}.
\end{eqnarray}
In what follows, we shall consider spherically symmetric space--times with line element parametrized as
\begin{eqnarray}\label{le}
ds^{2}=-e^{\nu}dt^{2}+e^{\lambda}dr^{2}+r^{2}d\Omega^{2},
\end{eqnarray}
where $\nu$ and $\lambda$ are functions of the radial coordinate $r$ only and matter sector given by $T^{\mu}_{\nu}=diag(-\rho,p_{r},p_{\perp},p_{\perp})$ and $\theta^{\mu}_{\nu}=diag(-\rho^{\theta},p_{r}^{\theta},p_{\perp}^{\theta},p_{\perp}^{\theta})$. It should be noted that, since Einstein's tensor is divergence free, the total energy--momentum tensor $T^{(tot)}_{\mu\nu}$
satisfies
\begin{eqnarray}\label{cons}
\nabla_{\mu}T^{(tot)}_{\mu\nu}=0.
\end{eqnarray}
Now, considering Eq. (\ref{le}) as a solution of the Einstein equations (\ref{eq5}) and using (\ref{total}), we obtain
\begin{eqnarray}
\kappa^{2} \tilde{\rho}&=&\frac{1}{r^{2}}+e^{-\lambda}\left(\frac{\lambda'}{r}-\frac{1}{r^{2}}\right)\label{eins1}\\
\kappa^{2} \tilde{p}_{r}&=&-\frac{1}{r^{2}}+e^{-\lambda}\left(\frac{\nu'}{r}+\frac{1}{r^{2}}\right)\label{eins2}\\
\kappa^{2} \tilde{p}_{\perp}&=&\frac{e^{-\lambda}}{4}\left(\nu'^{2}-\nu '\lambda '+2\nu''
+2\frac{\nu'-\lambda'}{r}\right)\label{eins3},
\end{eqnarray}
where the primes denote derivation with respect to the radial coordinate and we have defined
\begin{eqnarray}
\tilde{\rho}&=&\rho+\alpha\rho^{\theta}\label{rot}\\
\tilde{p}_{r}&=&p_{r}+\alpha p_{r}^{\theta}\label{prt}\\
\tilde{p}_{\perp}&=&p_{r}+\alpha p_{\perp}^{\theta}.\label{ppt}
\end{eqnarray} 
It is worth mentioning that decomposition (\ref{total}) seems like a na\"{i}ve separation of the constituents of the matter content in the sense that, given the non--linearity of Einstein's equations, the resulting system  does not correspond to  a set of equation for each source involved. However, contrary to what is commonly believed, the decoupling of the equations can be successfully  achieved in the framework of MGD because the separation of the total energy momentum tensor is complemented with a geometric deformation in the metric functions given by
\begin{eqnarray}
\nu&=&\xi+\alpha g\\
e^{-\lambda}&=&\mu +\alpha f\label{def}.
\end{eqnarray}
In the above expression, $\{g,f\}$ are the so--called decoupling functions and $\alpha$ is the decoupling parameter appearing in Eq. (\ref{efq})
which can be alternatively interpreted as the quantity 
that controls the geometric deformation undergone by the metric components, $\{\xi,\mu\}$. Although a general treatment considering deformation in both components of the metric is possible (see Ref. \cite{ovalleplb}), in this work we  will concentrate into the particular case $g=0$ and $f\ne 0$, namely, we will only deform the radial component of the metric, $g^{rr}$, which corresponds to the case of minimal deformation. By doing so, we obtain
two sets of differential equations:
the set
\begin{eqnarray}
\kappa ^2 \rho &=&\frac{1-r \mu'-\mu}{r^{2}}\label{iso1}\\
\kappa ^2 p_{r}&=&\frac{r \mu  \nu '+\mu -1}{r^{2}}\label{iso2}\\
\kappa ^2 p_{\perp}&=&\frac{\mu ' \left(r \nu '+2\right)+\mu  \left(2 r \nu ''
+r \nu '^2+2 \nu '\right)}{4 r}\label{iso3},
\end{eqnarray}
with
\begin{eqnarray}
\nabla_{\mu}T^{\mu}_{\nu}=(p_{r})'
+\frac{\nu'}{2}(\rho+p_{r})-\frac{2}{r}(
p_{\perp}-p_{r})=0,
\end{eqnarray}{}
for $T_{\mu\nu}$
and
\begin{eqnarray}
\kappa ^2  \rho^{\theta}&=&-\frac{r f'+f}{r^{2}}\label{aniso1}\\
\kappa ^2 p_{r}^{\theta}&=&\frac{r f \nu '+f}{r^{2}}\label{aniso2}\\
\kappa ^2p_{\perp}^{\theta}&=&\frac{f' \left(r \nu '+2\right)+f \left(2 r \nu ''+r \nu '^2+2 \nu '\right)}{4 r}\label{aniso3},
\end{eqnarray}
for $\theta_{\mu\nu}$. Note that whenever 
$p_{r}^{\theta}\ne p_{\perp}^{\theta}$, the decoupling matter allows to induce an extra local anisotropy in the system. Now, the conservation equation, $\nabla_{\mu}\theta^{\mu}_{\nu}=0$, leads to
\begin{eqnarray}\label{tovtheta}
(p_{r}^{\theta})'+\frac{\nu'}{2}(\rho^{\theta}+p_{r}^{\theta})
-\frac{2}{r}(p_{\perp}^{\theta}-p_{r}^{\theta})=0,
\end{eqnarray}
which is a linear combination of Eqs. (\ref{aniso1}), (\ref{aniso2}) and (\ref{aniso3}). In this
sense, there is not exchange of energy--momentum tensor between the fluids  so that the interaction is purely gravitational. What is more, 
given the way the sources interact, the $\theta$--sector can be thought as dark matter or any other source whose nature could be unknown. It is remarkable that, although Eqs. (\ref{aniso1}), (\ref{aniso2})
and (\ref{aniso3}) are not Einstein's equations (there is a missing $1/r^{2}$ term), Eq. (\ref{tovtheta}) corresponds to the standard TOV equation.
\\

To complete the MGD program the next step consists in to provide $\{\nu,\mu\}$ 
(sourced by $\{\rho,p_{r},p_{\perp}\}$)  and to use (\ref{aniso1}), (\ref{aniso2}) and (\ref{aniso3})
in combination with a suitable complementary condition to obtain $f$. For example, when the method is employed to extend interior solutions embedded in a Schwarzschild vacuum, the continuity of the second fundamental form is ensured if the so called mimic constraint for the radial pressure \cite{ovalle2017}, namely,
\begin{eqnarray}\label{mimet}
-p_{r}^{\theta}= p,
\end{eqnarray}
is imposed as a complementary condition to close the system. Remarkably, the mimic constraint for the radial pressure leads to an algebraic equation for $f$ so that, in principle, any isotropic solution can be extended by using (\ref{mimet}). Another possibility is to  implement the mimic constraint for the density which leads to a differential equation for $f$, which can be solved in some situations (see for example \cite{ovalle2018}).

In the case of exterior solutions, the use of mimic constraints does not have any sense (or there is not any fundamental physical reason to consider them, at least). In this respect, the strategy consists in to impose some equations of state for the decoupling sector (barotropic, polytropic, or a linear combination relating the different components of $\theta_{\mu\nu}$) as in \cite{ovalle2018a,rincon19}. The other possibility is to impose suitable conditions, as for example, regularity in the TOV equation of the decoupling sector as we will illustrate in the next section.

\section{Regularity condition in the decoupling sector}\label{new}
As it is well known, the study of interior isotropic solutions require the setting of extra information  either  an  equation  of  state  or  a  geometrical  constraint to close the system. Similarly, when local anisotropy is introduced we need to impose an  additional condition which usually corresponds to a constraint on the anisotropy function. In the pioneering works by Bowers--Liang \cite{bowers} and Cosenza {\it et. al.}, \cite{cosenza1,cosenza2}  the additional condition is such that the TOV equation remains free of singularities, namely
\begin{eqnarray}
\tilde{p}_{\perp}-\tilde{p}_{r}=c\tilde{f}(\tilde{\rho},r)(\tilde{\rho}+\tilde{p}_{r})r^{n}.
\end{eqnarray}
In the above expression  $c$ and $n$ are constants and $\tilde{f}(\tilde{\rho},r)$ is, in principle, an arbitrary function which allows to regularize the TOV equation. In the present work, we adapt this strategy to deal with exterior solutions in the context of MGD as follows. First, note that 
as $T_{\mu\nu}$ corresponds to the matter sector of a well--known  solution, extra conditions are not required. However, the
$\theta$--sector must be constrained in order to close the system, so that we propose
\begin{eqnarray}\label{BLN}
p_{\perp}^{\theta}-p_{r}^{\theta}=c \tilde{f}(\theta^{1}_{1},r)(\rho^{\theta}+p_{r}^{\theta})r^{n},
\end{eqnarray}
and, as $\tilde{f}$ remains as an arbitrary function, we implement the Cosenza--Herrera--Esculpi--Witten ansatz which reads
\begin{eqnarray}\label{cosen}
\tilde{f}(\tilde{\rho},r)=r^{1-n}\,\nu'/2.
\end{eqnarray}
At this point a couple of comments are in order. First, it is worth mentioning that the same strategy was successfully implemented to extend the Tolman IV solution in Ref. \cite{Abellan2}. 
Second,  the regularity condition imposed in the TOV equation associated with the decoupling sector is not an obligated requirement. However, there are
no reasons to assume that the $\theta$--sector suffers from the same pathology that the original background we are extending by MGD.

Now, it is clear that the replacement of (\ref{aniso1}), (\ref{aniso2}) and (\ref{aniso3}) in (\ref{BLN}), leads to a differential equation for $f$ where the only required information is the metric function $\nu$ which, in the context of MGD, is common for the two sectors involved.
Indeed, from Eqs. (\ref{BLN}) and (\ref{cosen}) we obtain
\begin{eqnarray}
\left((2 c +1)r \nu '+2\right)f'+ \left( \left(r(1-2 c ) \nu '-2\right)\nu '+2 r \nu ''-\frac{4}{r}\right)f=0,
\end{eqnarray}
which can be formally solved to obtain
\begin{eqnarray}\label{diff}
f=c_1 e^{ \int \frac{u \left(\nu ' \left((2 c-1) u 
	\nu '+2\right)-2 u \nu ''\right)+4}{u \left((2 c+1) u \nu '+2\right)} \, du}.
\end{eqnarray}
It is worth noticing that finding an analytical solution of the above integral will depend on the particular form of the metric function $\nu$. In this work, we shall consider the Schwarzschild black hole as a seed solution of the method to illustrate how it works. 

\subsection{Schwarzschild solution}
In this particular case the seed space--time corresponds to
\begin{eqnarray}\label{Sch}
e^{\nu}&=&e^{-\lambda}=1-\frac{2M}{r}.
\end{eqnarray}
Next, replacing (\ref{Sch}) in (\ref{diff}) we obtain
\begin{eqnarray}
f=c_{1} \bigg(1-\frac{2 M}{r}\bigg) \bigg((2 c-1) M+r\bigg)^2,
\end{eqnarray}
which determines the decoupling sector. Note that $c_{1}$ should be a constant with dimensions of the inverse of length squared. Now, from Eq. (\ref{def}), the $g^{rr}$ component of the metric reads
\begin{eqnarray}\label{deform}
e^{-\lambda}=\bigg(1-\frac{2 M}{r}\bigg)\bigg(1+\alpha c_{1}  \bigg((2 c-1) M+r\bigg)^2\bigg).
\end{eqnarray}
It is worth noticing that the solution is not asymptotically flat unless $\alpha\to0$. Indeed, after expanding (\ref{deform}) we obtain
\begin{eqnarray}
e^{-\lambda}=A+ \frac{B}{r}+ 
C r+D r^{2},
\end{eqnarray}
with 
\begin{eqnarray}
A&=&\alpha  (2 c-1)^2 c_{1} M^2-4 \alpha  (2 c-1) c_{1} M^2+1\\
B&=&-2 \alpha  (2 c-1)^2 c_{1} M^3-2 M\\
C&=&2 \alpha  (2 c-1) c_{1} M-2 \alpha  c_{1} M\\
D&=&\alpha  c_{1}.
\end{eqnarray}
In this sense, an asymptotically flat solution is obtained whenever $C=D=0$
which can be achieved
when $\alpha=0$. Another possibility is to take
$c=1$ which leads to 
\begin{eqnarray}
e^{-\lambda}=1
-3 \alpha  c_{1} M^2
+\frac{-2M}{r}(1+\alpha c_{1}M^{2})+
\alpha  c_{1} r^2.
\end{eqnarray}
It should be noted that, when $\alpha c_{1}<<\frac{1}{M^{2}}$ and $r\to\infty$ the metric of the total solution reads
\begin{eqnarray}\label{m3}
ds^{2}\approx-dt^{2}+\frac{dr^{2}}{1+\alpha c_{1}r^{2}}+r^{2}d\Omega^{2}.
\end{eqnarray}
Interestingly, after the coordinate transformation $\alpha c_{1} r^{2}=-k\tilde{r}^{2}$,  the induced 3--metric in (\ref{m3}) can be written as
\begin{eqnarray}\label{3}
ds^{2}_{(3)}=R^{2}\left(\frac{d\tilde{r}^{2}}{1-k \tilde{r}^{2}}+\tilde{r}^{2}d\Omega^{2}\right),
\end{eqnarray}
with $R^{2}=-\frac{k}{\alpha c_{1}}$ and $R\in\mathbb{R}$.
As can be seen, the above metric describes a manifold with constant curvature \cite{podolski}
\begin{eqnarray}
\mathcal{K}=\frac{k}{R},
\end{eqnarray}
with $k$ the sign of the curvature.
Furthermore, as it is well known, the sign of $k$ allows  distinguishing three possibilities \cite{podolski}
\begin{enumerate}
    \item $k=0$, corresponds to a flat space or Euclidean 3--space $E^{3}$. Of course, this is equivalent to setting
$\alpha=0$ in which case the solution reduces to the Schwarzschild geometry.
    \item $k=1$, represents a 3--space of constant curvature with topology $S^{3}$. Indeed, by introducing $r=\sin\chi$, the metric (\ref{3}) reads
    \begin{eqnarray}
    ds^{2}_{(3)}=R^{2}(d\chi^{2}+\sin^{2}\chi(d\theta^{2}+\sin^{2}\theta d\phi^{2})).
    \end{eqnarray}
Note that, in order to ensure $R\in\mathbb{R}$, we need to impose
$\alpha c_{1}<0$. Besides,
it is worth mentioning that this case corresponds to the Einstein's static universe model.
\item $k=-1$, corresponds to a Lobatchevski space of constant negative curvature or a hyperbolic 3--space $H^{3}$. In this case, after defining  $r=\sinh{\chi}$, we have
\begin{eqnarray}
ds^{2}_{(3)}=R^{2}
(d\chi^{2}+\sinh^{2}\chi(d\theta^{2}+\sin^{2}\theta d\phi^{2})).
\end{eqnarray}
In this case $R\in\mathbb{R}$ whenever
$\alpha c_{1}>0$.
\end{enumerate}
It is remarkable how the physics of the asymptotic geometry
of the MGD--extended solution depends on the nature of the decoupling parameter $\alpha$. In this sense, we can interpret the gravitational decoupling as a map allowing topology changes of the space--like slices from $\mathbb{R}\times S^{2}$ (the Schwarzschild case) to $E^{3}$, $S^{3}$, or $H^{3}$ depending on the sign of $\alpha c_{1}$. Our results are summarized in table \ref{table1}.
\begin{table}[h!]
    \centering
    \begin{tabular}{||c |c ||} 
\hline
  Topology change&  $\alpha c_{1}$ \\
\hline
$\mathbb{R}\times S^{2}\ \to\ E^{3}$&$0$\\
\hline
$\mathbb{R}\times S^{2}\ \to\ S^{3}$&$<0$\\
\hline
$\mathbb{R}\times S^{2}\ \to\ H^{3}$& $>0$\\
\hline
    \end{tabular}
    \caption{Asymptotic behaviour of the exterior solution.}
    \label{table1}
\end{table}

We would like to remark that, although the MGD--extended solution obtained here can not be interpreted as an exterior  geometry because  it is neither asymptotically flat nor asymptotically anti--de Sitter, as required,
the induced topological change is an interesting fact. Indeed, the asymptotic behavior described in case (ii) resembles the Einstein--Straus (ES) model \cite{es} in which
the asymptotically flat exterior of a Schwarzschild space--time is replaced by a section of a Friedman universe, continuously matched with the remaining interior of the Schwarzschild solution \cite{susman}. 
It is worth mentioning that some authors \cite{s2,s3,s4} have considered this simple hybrid space--time,
particularly in its application to cosmology as a simple model for
inhomogeneities in a Friedman background. Now, such a resemblance must be considered as formal
for two main reasons. First, the asymptotic region obtained here corresponds to
the Einstein universe in contrast to the Friedman space
in the ES model. Second, instead of using  Darmois \cite{darmois} conditions to match the interior and exterior geometries as in the ES case, we are using the gravitational decoupling to bring the matching in a straightforward manner.

In what follows, we 
will study some geometrical properties of the MGD--extended solution. 
First, note that 
for $\alpha=0$
 the event horizon of the original solution is located at $r_{H}=2M$. Now, as
 the signs of the metric components $g_{tt}$ and $g_{rr}$ get exchanged across $r=r_{H}$ the apparition of extra roots of $e^{-\lambda}$ for $r\ne r_{H}$ would lead to metrics with signatures $(--++)$ which must be discarded. In this particular case, non--extra roots appear for every $r\ne r_{H} $, whenever the equation 
\begin{eqnarray}\label{const}
1+\alpha c_{1}  \bigg((2 c-1) M+r\bigg)^2=0,
\end{eqnarray}
leads to either $r<0$ or $r\in \mathbb{C}$. Now, a straightforward computation reveals that Eq. (\ref{const}) yields 
\begin{eqnarray}
r_{1}&=&\frac{-2i\sqrt{c_{1}\alpha}+2c_{1}M\alpha
	-4 c c_{1}M\alpha}{2c_{1}\alpha}\label{r1}\\
r_{2}&=&\frac{2i\sqrt{c_{1}\alpha}+2c_{1}M\alpha-4c c_{1}M\alpha}{2c_{1}\alpha}\label{r2}.
\end{eqnarray}
A first possibility is to choose $\alpha c_{1}>0$ which entails $r\in \mathbb{C}$. The second possibility is to take $\alpha c_{1}<0$ from where
Eqs. (\ref{r1}) and (\ref{r2}) read
\begin{eqnarray}\label{eq33}
r_{1}&=&-\frac{2\sqrt{|c_{1}\alpha|}-2|c_{1}\alpha|M
	+4 c |c_{1}\alpha|M}{2|c_{1}\alpha|} \\ \label{eq34}
r_{2}&=&-\frac{-2\sqrt{|c_{1}\alpha|}
	-2|c_{1}\alpha|M+4c |c_{1}\alpha|M}{2|c_{1}\alpha|}.
\end{eqnarray}
Now, to enforce $r_{1},r_{2}<0$ we need to impose
\begin{eqnarray}
2\sqrt{|c_{1}\alpha|}-2|c_{1}\alpha|M
+4 c |c_{1}\alpha|M>0\\
-2\sqrt{|c_{1}\alpha|}
-2|c_{1}\alpha|M+4c |c_{1}\alpha|M>0,
\end{eqnarray}
from where we obtain $c>\frac{1}{2}$ and
\begin{eqnarray}
|c_{1}\alpha|>\frac{1}{M^{2}-4c M^{2}+4c^{2}M^{2}}.
\end{eqnarray}
 
It is worth mentioning that, the imposition of regularity on the TOV equation by the Cosenza--Herrera--Esculpi--Witten anisitropy (see Eqs. (\ref{BLN}) and (\ref{cosen})), leads to an extended Schwarzschild solution that maintains the same singular point at $r=0$ and the apparition of extra singularities is impossible for every value of the involved constants. Indeed, the Ricci, $R$, Ricci squared, $\mathcal{R}$, and Kretschmann scalar ,$K$, read
\begin{equation}\label{Ricci}
R=-\frac{\alpha c_{1}\left[2\left(2-5c+2c^{2}\right)M^{2}+\left(8c-7\right)Mr+3r^{2}\right]}{r^{2}},
\end{equation}
\begin{equation}\label{RicciSquare}
\mathcal{R}=\frac{4\alpha^{2}c^{2}_{1}}{r^{4}}\bigg[\left(2c-1\right)M+r\bigg]^{2}\bigg[\left(2c^{2}-6c+7\right)M^{2} 
\left(4c-9\right)Mr+3r^{2}\bigg],   
\end{equation}
\begin{eqnarray}\label{Kretsch}
K&=&2\alpha\left(2c-1\right) c_{1}M^{3}r\big(20+27c_{1}\alpha r^{2}+8\alpha c^{2}c_{1}r^{2}
-4c\left(1+7\alpha c_{1}r^{2}\right)\big)\bigg]\nonumber\\
&&+3\alpha^{2}c^{2}_{1}r^{6}+16\alpha^{2}\left(c-1\right)c^{2}_{1}Mr^{5}\nonumber\\
&&+2\alpha\left(1-2c^{2}\right)^{2}c_{1}M^{4}\left(12+\alpha\left(2c^{2}-14c+29\right)c_{1}M^{2}\right)\nonumber\\
&&+M^{2}\big(12-8\alpha\left(c-2\right) c_{1}r^{2}+\alpha^{2}\left(32c^{2}-72c+37\right)c^{2}_{1}r^{4}\big)\nonumber\\
&&+\frac{4}{r^{6}}\bigg[12\alpha^{2}\left(1-2c\right)^{4}c^{2}_{1}M^{6}-8\alpha^{2}\left(c-5\right)
\left(2c-1\right)^{3}c^{2}_{1}M^{5}r.
\end{eqnarray}
Note that, when $\alpha$ is set to zero Eqs. (\ref{Ricci}), (\ref{RicciSquare}), (\ref{Kretsch}) correspond to
the scalars of the Schwarzschild solution, as expected.

\begin{figure*}[ht]
\centering
\includegraphics[width=0.45\textwidth]{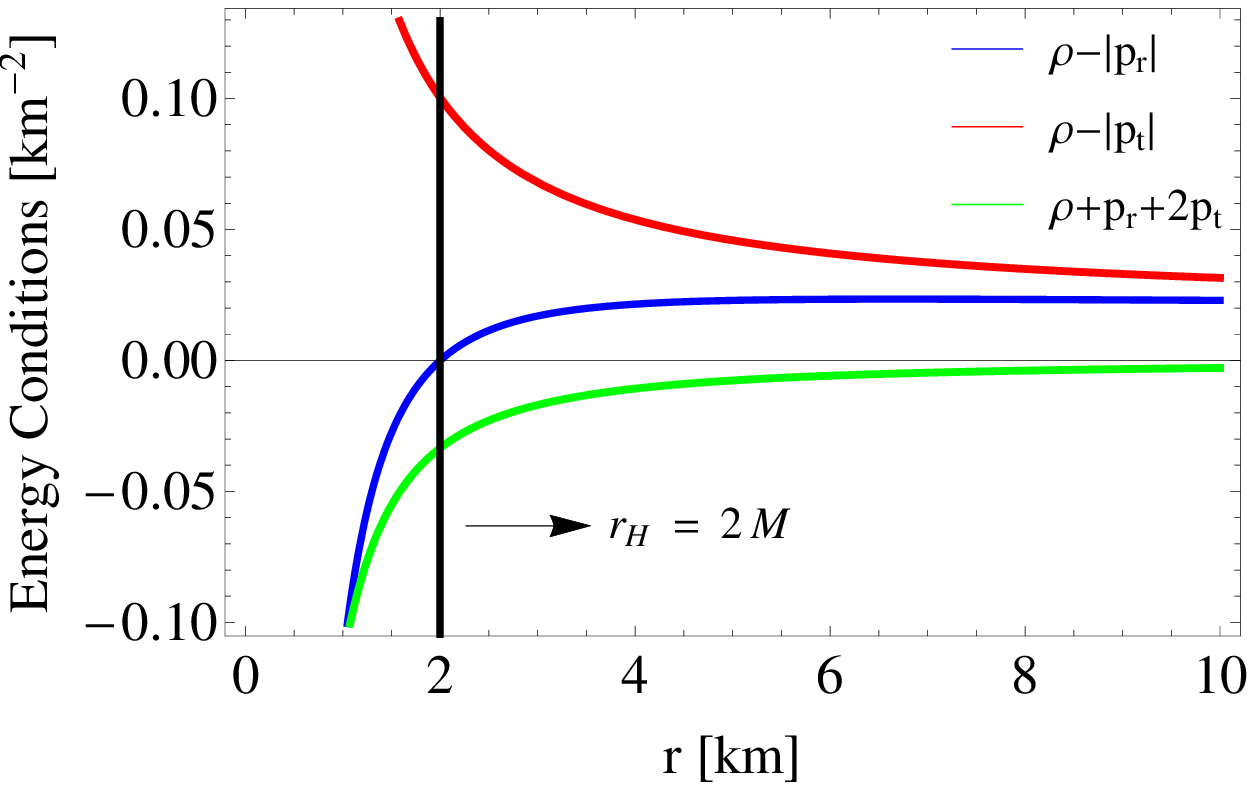}     \
\includegraphics[width=0.45\textwidth]{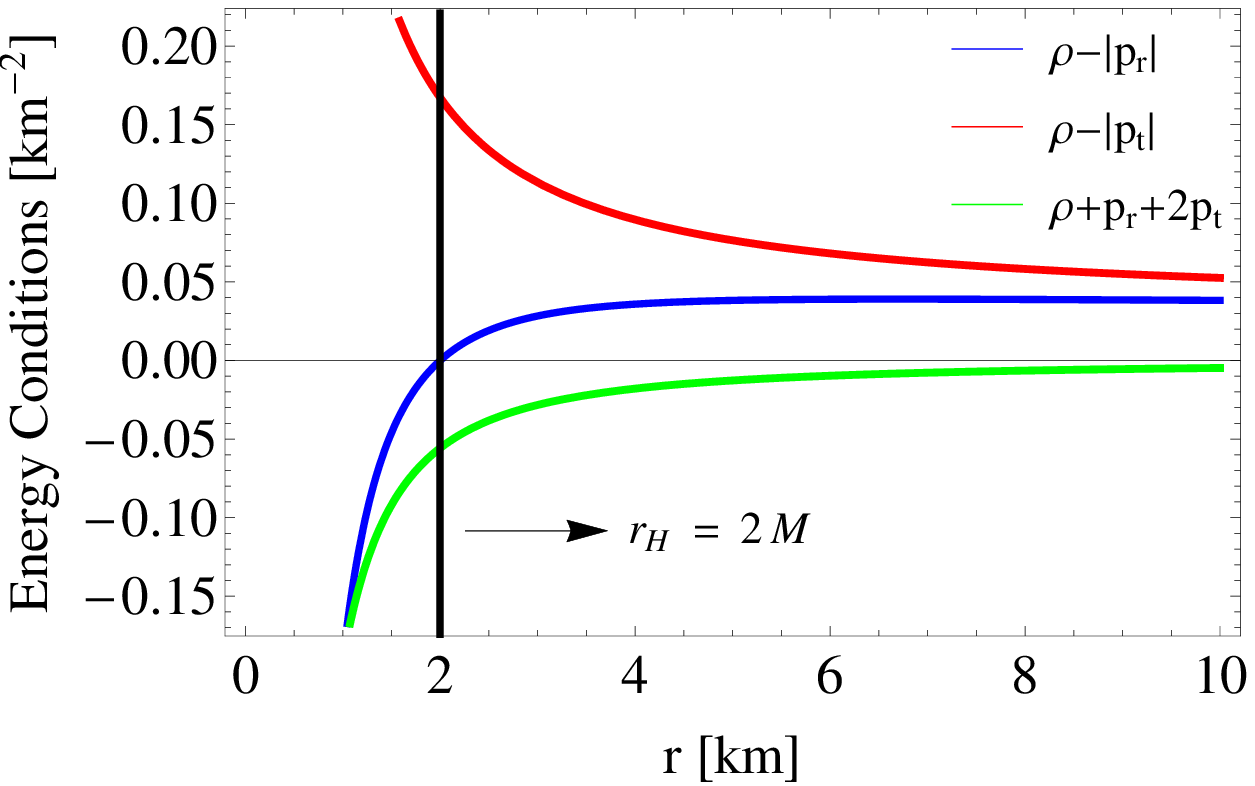}      \
\caption{
The null, weak and strong energy conditions against the radial coordinate $r$ for $M=1$, $c=3$ and $\{\alpha=0.2;c_{1}=-1.2\}$ (left panel) and $\{\alpha=-0.5;c_{1}=0.8\}$ (right panel).
}
\label{fig1}
\end{figure*}

From now on, we shall focus our attention on the study of the matter sector sustaining the extended solution. From (\ref{eins1}), (\ref{eins2}) and (\ref{eins3}) we obtain
\begin{eqnarray}
\tilde{\rho}&=&-\frac{\alpha  c_{1} ((2 c-1) M+r) ((2 c-5) M+3 r)}{8 \pi  r^2}\label{til1}\\
\tilde{p}_{r}&=&\frac{\alpha  c_{1} ((2 c-1) M+r)^2}{8 \pi  r^2}\label{til2}\\
\tilde{p}_{\perp}&=&-\frac{\alpha  c_{1} (M-r) ((2 c-1) M+r)}{8 \pi  r^2}\label{til3}.
\end{eqnarray}
Now, from Eq. (\ref{til1}) it is clear that the solution with $\alpha c_{1}>0$ (asymptotically $H^{3}$) must be discarded to
ensure the positivity of the energy density. Besides, in order to avoid real roots of $\tilde{\rho}$
we need to restrict the parameter $c$, namely, $c>\frac{5}{2}$.\\
Let us now consider the energy conditions. The null energy condition states that
\begin{eqnarray}
\tilde{\rho}+\tilde{p}_{r}&\ge& 0\\
\tilde{\rho}+\tilde{p}_{\perp}&\ge& 0,
\end{eqnarray}
which leads to
\begin{eqnarray}
\tilde{\rho}+\tilde{p}_{r}&=&\frac{2c_{1}\alpha(2M-r)
((2c-1)M+r)}{8\pi r^{2}}\ge 0\label{n1}\\
\tilde{\rho}+\tilde{p}_{\perp}&=&
-\frac{c_{1}\alpha((c-2)M+r)((2c-1)M+r)}{4\pi r^{2}}\ge 0\label{n2}.
\end{eqnarray}
It is noticeable that, expression (\ref{n2}) is satisfied everywhere whenever $c>2$. However, Eq. (\ref{n1}) can be fulfilled for $r>2M$ only (out the event horizon). Now, the above condition combined with the constraint for $\tilde{\rho}>0$ corresponds to the weak energy condition (WEC). In this sense we say that our solution satisfies the WEC out the event horizon.
The dominant energy condition (DEC) states that
\begin{eqnarray}
\tilde{\rho}-|\tilde{p}_{r}|\ge 0\label{d1}\\
\tilde{\rho}-|\tilde{p}_{\perp}| \ge 0.\label{d2}
\end{eqnarray}
As can be seen from Eq. (\ref{til2}), $\tilde{p}_{r}$ is negative everywhere, so 
Eq. (\ref{d1}) leads to $\tilde{\rho}+\tilde{p}_{r}$ which coincides with (\ref{n1}). In other words, the requirement given by (\ref{d1}) is satisfied out the event horizon only. Regarding Eq. (\ref{d2}) the situation is subtler. Indeed, Eq. (\ref{til3}) reveals that $\tilde{p}_{\perp}>0$ for $r<M$ an $\tilde{p}_{\perp}<0$ for $r>M$. In this manner, for
$r<M$, Eq. (\ref{d2}) reads
\begin{eqnarray}
\tilde{\rho}-\tilde{p}_{\perp}=-\frac{\alpha  c_{1} ((2 c-1) M+r) ((c-3) M+2 r)}{4 \pi  r^2},
\end{eqnarray}
which is positive when $c>3$. Similarly, for $r>M$ we have that (\ref{d2}) reduces to $\tilde{\rho}+\tilde{p}_{\perp}>0$, which coincides with (\ref{n2}) and it is fulfilled everywhere for $c>2$. Finally we study the strong energy condition (SEC) that states 
\begin{eqnarray}
\tilde{\rho}+\tilde{p}_{r}+2p_{\perp}\ge 0,
\end{eqnarray}
in combination with Eqs. (\ref{n1}) and (\ref{n2}) (which are satisfied out the event horizon). Now, from Eqs. (\ref{til1}), (\ref{til2}) and (\ref{til3}) we have
\begin{eqnarray}
\tilde{\rho}+\tilde{p}_{r}+2p_{\perp}=\frac{\alpha  c_{1} M ((2 c-1) M+r)}{4 \pi  r^2},
\end{eqnarray}
which is negative everywhere for the accepted values of $c$. In summary, we have found that the conditions of regularity of the TOV equation of the decoupling sector, lead to an extension of the Schwarszchild exterior solution, which satisfies the DEC out the event horizon but violates the SEC everywhere. The previous discussion about the energy conditions is corroborated in Fig. \ref{fig1} which displays the behaviour of these constraints on the components of the energy momentum tensor versus the radial coordinate $r$. 

\section{Final remarks}\label{remarks}
In this work, we implemented the
gravitational decoupling in the framework of the Minimal Geometric Deformation method to study its effect on exterior solutions by imposing a regularity condition in the Tolman-–Oppenheimer-–Volkoff equation. More precisely, we followed the pioneering work by Bowers and Liang to propose a particular form of the anisotropy function on the decoupling sector of the solution. We obtained that the decoupling function, $f$, admits a formal solution in terms of an integral involving the metric potential associated  with the $g_{tt}$
component of the metric. Based on this
result, we conclude that, although it is possible  to extend any exterior solution with the methodology here developed, obtaining an analytical solution for $f$ depends on the particular form of the seed. To illustrate the method, we used the Schwarzschild exterior as a seed and obtained that, after a detailed analysis of the involved parameters, the minimally deformed geometry is asymptotically connected to
a constant curvature manifold
which character depends on the sign of the decoupling parameter. In this sense, we can interpret the Minimal Geometric Deformation method as a kind of mechanism that allows the change of topologies from $\mathbb{R}\times S^{2}$ when the decoupling parameter is turned off to either $E^{3}$, $S^{3}$ or $H^{3}$ depending on the sign of $\alpha$. Besides, we obtained that the solution satisfies the dominant energy condition out the even horizon and violates the strong energy condition everywhere.

\section*{Acknowledgements}
F. Tello-Ortiz thanks the financial support by the CONICYT PFCHA/DOCTORADO-NACIONAL/2019-21190856 and projects ANT-1856 and SEM 18-02 at the Universidad de Antofagasta, Chile, and grant Fondecyt No. $1161192$, Chile. S. K. Maurya and F. Tello-Ortiz thank to TRC project-BFP/RGP/CBS/19/099 of the Sultanate of Oman.

\end{document}